\documentclass[prl,twocolumn,amsmath,amssymb]{revtex4}
\usepackage{amsfonts}
\usepackage{amssymb}
\usepackage{graphicx}
\usepackage{subfigure}
\usepackage{color}
\usepackage{ulem}
\usepackage{bbold}
\usepackage{placeins}
\usepackage{soul,xcolor}
\usepackage{epstopdf}
\usepackage{bm}
\usepackage{braket}

\begin{document}

	\title{Leggett-Garg inequality in the context of three flavour neutrino oscillation}

	\author{Javid Naikoo}
	\email{naikoo.1@iitj.ac.in}
	\affiliation{Indian Institute of Technology Jodhpur, Jodhpur 342011, India}
	
	\author{Ashutosh Kumar Alok}
	\email{akalok@iitj.ac.in}
	\affiliation{Indian Institute of Technology Jodhpur, Jodhpur 342011, India}
	
	\author{Subhashish Banerjee}
	\email{subhashish@iitj.ac.in}
	\affiliation{Indian Institute of Technology Jodhpur, Jodhpur 342011, India}

	\author{S. Uma Sankar}
	\email{uma@phy.iitb.ac.in}
	\affiliation{Indian Institute of Technology Bombay, Powai 400076, India}

	\date{\today}
	
	\begin{abstract}
The present work is devoted to the characterization of the Leggett-Garg inequality  for three-flavoured neutrino oscillations in presence of both matter and Charge-Conjugation and Parity violating ($CP$) effects. This study complements and completes the recent one put forward in~\cite{Javid} by relaxing the stationary condition. At variance with the latter case, the LGI contains interference terms which cannot be expressed in terms of experimentally measurable quantities, thus drawing a clear-cut distinction between the two scenarios, as well as  highlighting the role of the stationary assumption on such systems. We find that the additional terms are small for high energy neutrino beam compared to the maximum value attained by the Leggett-Garg parameter.
	\end{abstract}

	\maketitle
	
	\section{Introduction}
	
One of the most characterizing aspects of quantum mechanics is the principle of superposition, according to which a system exists simultaneously in different states until a measurement is performed on it. A counterintuitive situation occurs when one tries to ascertain the quantum nature of mesoscopic systems, an endeavour that is broadly known as macroscopic quantum coherences (such as, e.g., the famous Schr\"{o}dinger's cat), which has been subjected to a number of investigations. Bell's inequality\cite{Bell} represents in this regard a milestone result, providing a quantitative criterion to distinguish between classical and quantum correlations between spatially separated systems.
In the temporal regime, Leggett-Garg inequalities (LGI) investigate the nature of correlations among measurements performed on the same system but at different instants of time. In \cite{Leggett}, the intuition about our view of macroscopic systems was formalized in terms of two principles, namely \textit{(i)} macroscopic realism (MR) and \textit{(ii)} non-invasive measurability (NIM)~\cite{Kofler,Fritz,Emary}. The former implies that measurements performed on the system of interest just reveal pre-existing values, while the latter (NIM) asserts that such measurements can be performed without changing the state of the system.
Beside providing a testbed for macroscopic coherences, LGI have thus been employed to investigate the notion of \textit{realism}, i.e., whether a description of the system under consideration can or cannot be given in terms of a set of hidden parameters. Any violation of LGI would automatically exclude any hidden-variable theory. 

 Quantum coherences extended over macroscopic distances have been in the spotlight in the context of flavour oscillations in neutrinos  and mesons \cite{KDqc,KDgp,JNtqc,JNelg,KDcm,AKAsin2beta,AKAqcBK}. This provides ample reason for  promoting such systems as candidates to study LGI.
Clearly, in this scenario the postulate of NIM makes contact with experiments somewhat difficult, usually leading to resort on the additional assumption of \textit{stationarity}, implying that correlations between different measurements only depend on time differences instead of specific time instants. 
This leads to a modified version of the LGI, called Leggett-Garg type Inequalities (LGtI), where all the intermediate non-measurable correlations are replaced by measurable ones \cite{Huelga1995, Huelga1996, Huelga1997, Waldher}. Such approach was recently used in ~\cite{Javid}, where the resulting LGtI can be recast in terms of experimentally measurable quantities such as neutrino flavour oscillation probabilities.

In the present work, however, we intend to complement and complete that analysis by considering the full LGI in its original setting and show that they are violated in the context of three flavour neutrino oscillations and matter interactions.
It is worth mentioning that various quantum information theoretic quantities  were considered in the context of two and three flavour neutrino oscillations also in ~\cite{2flav,3flav}. The factorization of the Hilbert-space was there achieved by the occupation number of neutrinos introduced in ~\cite{blasone}. Here, we study the LGI in the context of three flavour neutrino oscillations. A study of LG type inequlaity in the two and three flavour scenario was presented in \cite{Qiang} and \cite{Gango}, respectively.\par
In this contribution we investigate the full three flavour neutrino dynamics, which allows the inclusion of  $CP$ violations as well as the disentangling of the mass hierarchy problem. The paper is organized by introducing LG inequalities and the dynamics of neutrinos in vacuum and matter. Then we present the relevant correlators and analyse them generally. We make use of the experimental input parameters like \textit{energy} and \textit{baseline} of  the two ongoing experiments NO$\nu$A (NuMI Of-axis $\nu_e$ Appearance)\cite{Patterson,Adamson} and T2K(Tokai to Kamioka)~\cite{Abe} and the future experiment  DUNE (Deep Underground Neutrino Experiment)\cite{Dune}.

	\section{Leggett Garg inequalities in three flavour neutrino oscillations}
	
We first generally introduce  the Leggett Garg inequality and specify which one we will investigate. The next subsection discusses the time evolution of neutrinos in vacuum and matter, followed by the discussion of the Leggett Garg inequalities for neutrinos with a focus on the two ongoing experiments NO$\nu$A and T2K and also the future experiment DUNE.

	\subsection{Leggett Garg inequalities}
	
	Consider a quantum system with an underlying Hilbert space $\mathcal{H}$ and dynamics generated by an Hamiltonian H. Let moreover $\hat{Q}$ be a generic dichotomic observable (with possible outcomes $\pm 1$) satisfying the properties
	$\hat{Q}^{\dagger} = \hat{Q},\, \hat{Q}^{2} = \mathbb{1}$.
The two-time correlation function between the measurement of $\hat{Q}$ at times $t_i$ and $t_j$ ($t_i\geq t_j$) is given by the quantity $ C_{ij} \equiv \langle \hat{Q}(t_{i}) \hat{Q}(t_{j})\rangle$, where $\langle \ldots \rangle$ indicates the average performed over many repetitions. Here $\hat{Q}(t)$ denotes the time-evolution of the observable $\hat{Q}$ in Heisenberg picture, i.e. $\hat{Q}(t) \equiv \hat{U}^{\dagger}(t) \hat{Q} \hat{U}(t)$.
For a generic set of $n$ measurements of the dichotomic observable, the LGI provides a clear-cut bound on the parameter $K_n \equiv \sum_{i=1}^{n-1} C_{i,i+1} - C_{1,n}$ which allow to determine the existence of an hidden-variable theory describing the system of interest: for any $n\geq 3$, if \textit{realism} and \textit{non-invasive measurability} (NIM) are satisfied, then $K_n \leq n-2$.

In what follows we will focus on the first of such figures of merit, namely the LG parameter $K_3$ and its LG inequality
	\begin{equation} \label{K3-def}
	K_{3} = C_{01} + C_{12} - C_{02} \le 1
	\end{equation}
	whose violation would provide evidence that a realistic description of the system cannot be given. Quantum mechanical bound for $K_3$ is $\frac{3}{2}$ for a two level system \cite{Leggett}. It was shown in \cite{Budroni} that this bound holds for systems with arbitrary (but finite) number of levels, as long as the measurements are given by just two projectors \cite{Lambert, George, Kofler2007, Wilde}. In the limit $N \rightarrow \infty$, LGI can be violated up to its maximum algebraic sum \cite{Budroni2014}.\\
The simple form of Eq. (\ref{K3-def}) is the common feature of Bell type inequalities which are based on the usual Kolmogorovian rules of the probability. While the two quantities  $C_{01}$ and $C_{02}$ can always be easily expressed in terms of measurable quantities, since the first measurement of the observable $\hat{Q}$ occurs at the initial time $t_0$, the expression of the intermediate two-time correlation function  $C_{12}$ poses in general a real challenge since it depends on the whole history from the initial time $t_0$ to the second measurement time $t_1$.
	In order to bypass this difficulty, the NIM postulate is usually replaced by the weaker condition of \textit{stationarity}. Under the assumption that the time intervals $t_2 - t_1$ and $t_1 - t_0$ are equal to each other, the resulting LG type of Inequalities (LGtI) then take the much simpler form
	\begin{equation}\label{LGT}
	\tilde{K}_3 = 2 C_{01} - C_{02},
	\end{equation}
which again is bounded from above by one if realism is valid \cite{Huelga1995, Huelga1996}.	
    This LGtI was used in the context of neutrino oscillation \cite{Javid} to address the problem of neutrino mass hierarchy.\\
In this work we use LGI in its original setting and focus on the ongoing experimental facilities like NO$\nu$A and T2K as well as  the future DUNE experiment. The matter effect and CP violation are also taken into account in our analysis.
	
	\subsection{Neutrino state evolution in vacuum and  in constant matter density}
	The non zero mass square differences lead to the phenomena of neutrino oscillation, the existence of a flavour state $\ket{\nu_\alpha}$ into a coherent superposition of mass eigen states $\ket{\nu_k}$
	\begin{equation}\label{nu-alpha}
	\ket{\nu_\alpha} = \sum\limits_{k} U_{\alpha k}^* \ket{\nu_k}.
	\end{equation}
    where $U_{\alpha k}$ are the elements of a 3 $\times$ 3 unitary PMNS (Pontecorvo-Maki-Nakagawa-Sakata) mixing matrix $U$ parameterized by three mixing angles
    ($\theta_{12},\,\theta_{23},\,\theta_{13}$) and a $CP$ violating phase $\delta$. A convenient parametrization for $U(\theta_{12},\theta_{23},\theta_{32},\delta)$ is given by
    \begin{widetext}
    \begin{equation}\label{PMNS}
    U(\theta_{12},\theta_{23},\theta_{32},\delta) =
    \begin{pmatrix}
    c_{12} c_{13} & s_{12} c_{13} & s_{23} e^{-i \delta} \\ - s_{12}c_{23} - c_{12} s_{23}s_{13} e^{i\delta} & c_{12}c_{23}-s_{12}s_{23}s_{13} e^{i\delta} & s_{23}c_{13} \\ s_{13}s_{23} - c_{12}c_{23}s_{13} e^{i\delta} & -c_{12}s_{23}-s_{12}c_{23}s_{13} e^{i\delta} & c_{23}c_{13}\end{pmatrix}
    \end{equation}
    \end{widetext}
    where $c_{ij} = \cos\theta_{ij}$, $s_{ij} = \sin\theta_{ij}$,  $\theta_{ij}$ being the mixing angles and $\delta$ the $CP$ violating phase.  The experimental values for the PMNS mixing matrix are taken from the particle data group \cite{pdg}. Eq. (\ref{nu-alpha}) represents the state of the neutrino at time $t=0$. At a later time $t$, the mass eigen states are evolved according to the Schr\"odinger equation as
    \begin{align}
    \ket{\nu_\alpha (t)} &= \sum\limits_{k} U_{\alpha k}^* e^{-iE_k t} \ket{\nu_{k}}, \nonumber \\
                         &= \sum\limits_{\beta} \mathcal{A}_{\nu_\alpha \rightarrow \nu_\beta}(t) \ket{\nu_\beta},
    \end{align}
    where we have expanded $\ket{\nu_k}$ in the flavour basis $\ket{\nu_\beta}$ leading to the amplitudes $\mathcal{A}_{\nu_\alpha \rightarrow \nu_\beta}$, of transition from flavour $\nu_\alpha$ to $\nu_\beta$, given by
\begin{equation}
\mathcal{A}_{\nu_\alpha \rightarrow \nu_\beta}(t) = \sum\limits_{k} U_{\beta k} e^{-iE_k t} U_{\alpha k}^{*}.
\end{equation}
Consequently, the probability of transition at $t\approx L$ is given by
\begin{equation}
P_{\nu_\alpha \rightarrow \nu_\beta}(t) = | \mathcal{A}_{\nu_\alpha \rightarrow \nu_\beta}(t) |^2 = |\sum\limits_{k} U_{\beta k} e^{-iE_k t} U_{\alpha k}^{*}|^2.
\end{equation}
The amplitudes $\mathcal{A}_{\nu_\alpha \rightarrow \nu_\beta}(t)$ form the elements of the so called flavour evolution matrix $U_f(t)$. In matrix notation the state represented by the  vector $\bm{\nu}_{\alpha}(t) \equiv\left( \nu_e(t)\; \nu_\mu(t) \; \nu_\tau(t) \right)$ is connected to the state at $t=0$ by

\begin{equation}\label{ufvacuum}
 \bm{\nu}_{\alpha}(t)= U_{f}(t) \bm{\nu}(0).
\end{equation}

Neutrinos propagating through a constant matter density (with electron density $N_e$), characterized by the matter density parameter $A=\pm \sqrt{2} G_F N_e$,  interact weekly with the electrons in the medium. As a result of this interaction, the Hamiltonian $H_m = {\rm diag}[E_1, E_2, E_3]$ (in mass basis) picks up an interaction term $V_f = {\rm diag}[A, 0, 0]$ (in flavour basis). This leads to the following form of the flavour evolution matrix \cite{Tommy}
	\begin{equation} \label{ufmatter}
	U_{f}(L) = \phi ‎‎\sum_{n=1}^{3}  \frac{e^{-i \lambda_n L}}{3\lambda_{n}^{2} + c_1} \left[ (\lambda_{n}^{2}  + c_{1}) \mathbf{I} + \lambda_n \tilde{T} + \tilde{T}^2 \right],
	\end{equation}
	The phase  $\phi = e^{-i \frac{Tr\mathcal{H}_m}{3} L}$, $c_1 = det(T) tr(T^{-1})$ and the Hamiltonian in mass basis is $\mathcal{H}_m = H_m + U^{-1} V_f U$. The $\lambda_n$ are the eigenvalues of $T$ and the matrix $T$ and $\tilde{T}$ are given in~\cite{Tommy}. The flavour evolution operator, defined in Eq. (\ref{ufmatter}), can be used to deal with the situation when neutrinos pass through a series of matter densities with the matter density parameters  $A_1, A_2,\dots,A_n$. In this case, the total evolution operator becomes
	\begin{equation}
	U_{f}^{\rm tot}(L) = \prod_{i = 1}^{n} U_{f}(L_i).
	\end{equation}
	 Here, $L = ‎‎\sum_{i=1}^{n} L_i$ and $U_{f}(L_i)$ is evaluated for the density parameter $A_i$. A useful application of has been suggested for the mantle-core-mantle step function model simulating the Earth's matter density profile~\cite{TTommy}.
	\onecolumngrid
	\begin{widetext}
		\begin{figure}[ht]
		\centering
		\begin{tabular}{ccc}
			\includegraphics[width=60mm]{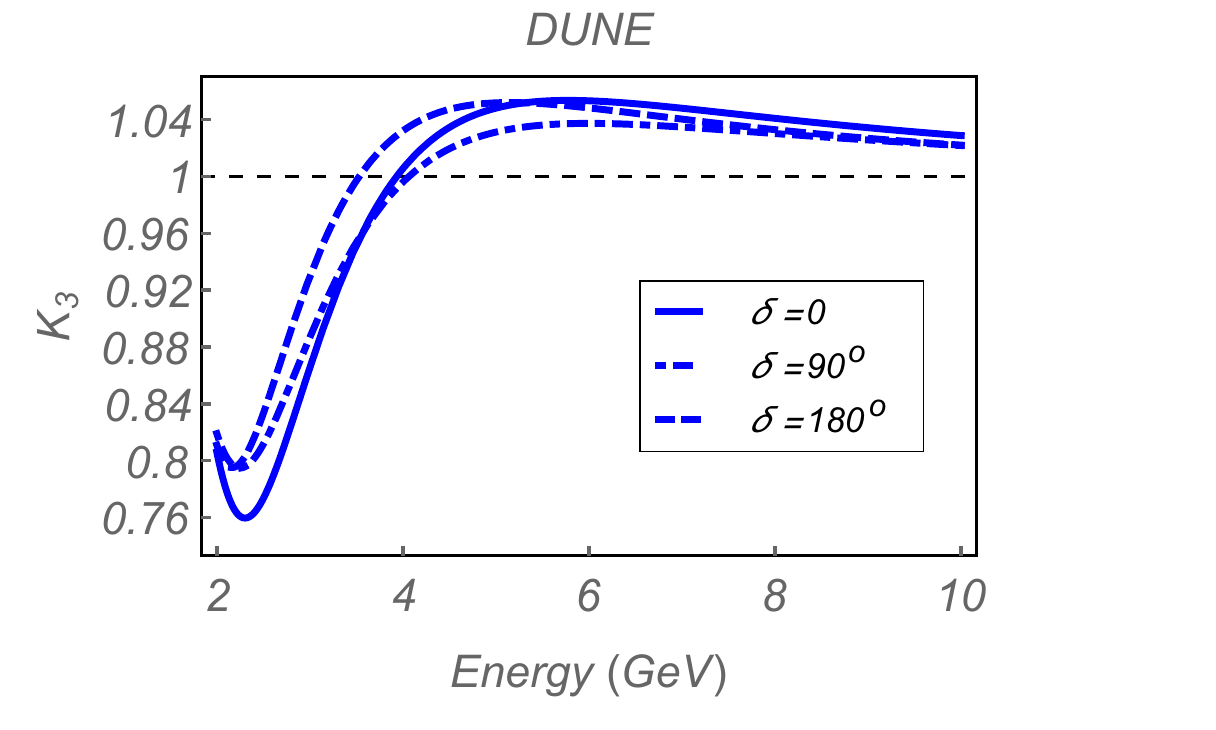}
			\includegraphics[width=60mm]{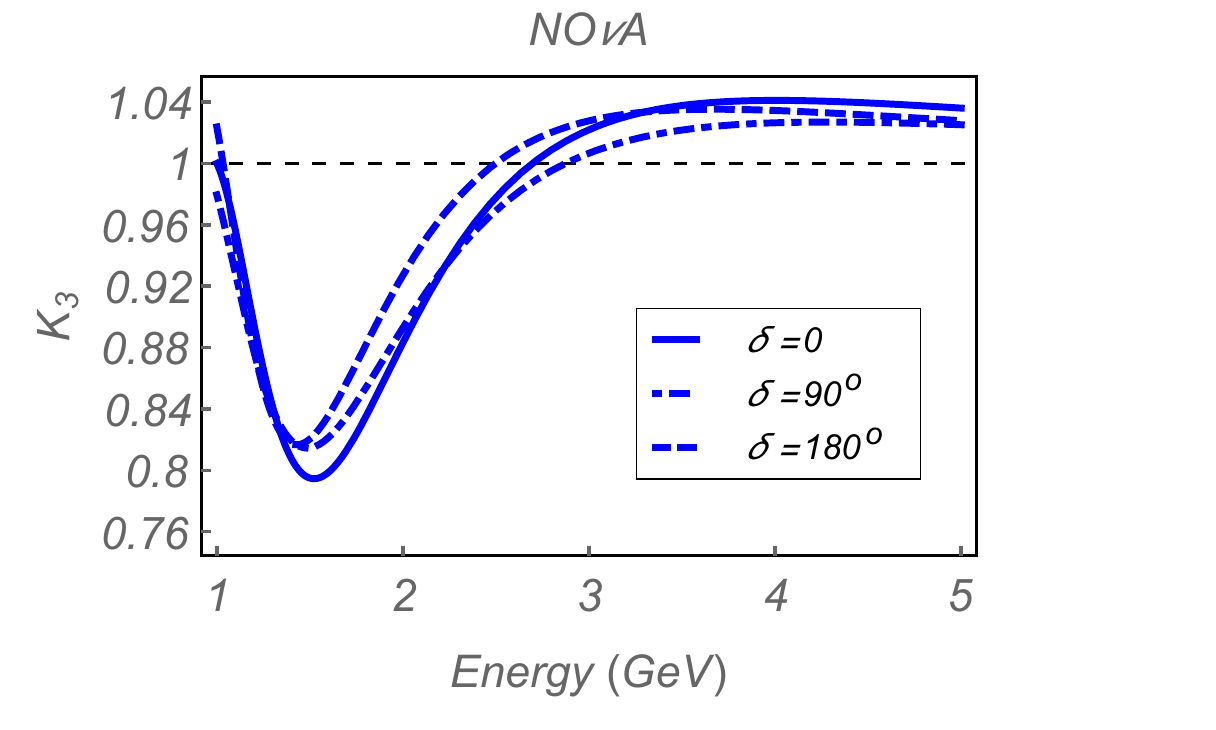}
			\includegraphics[width=60mm]{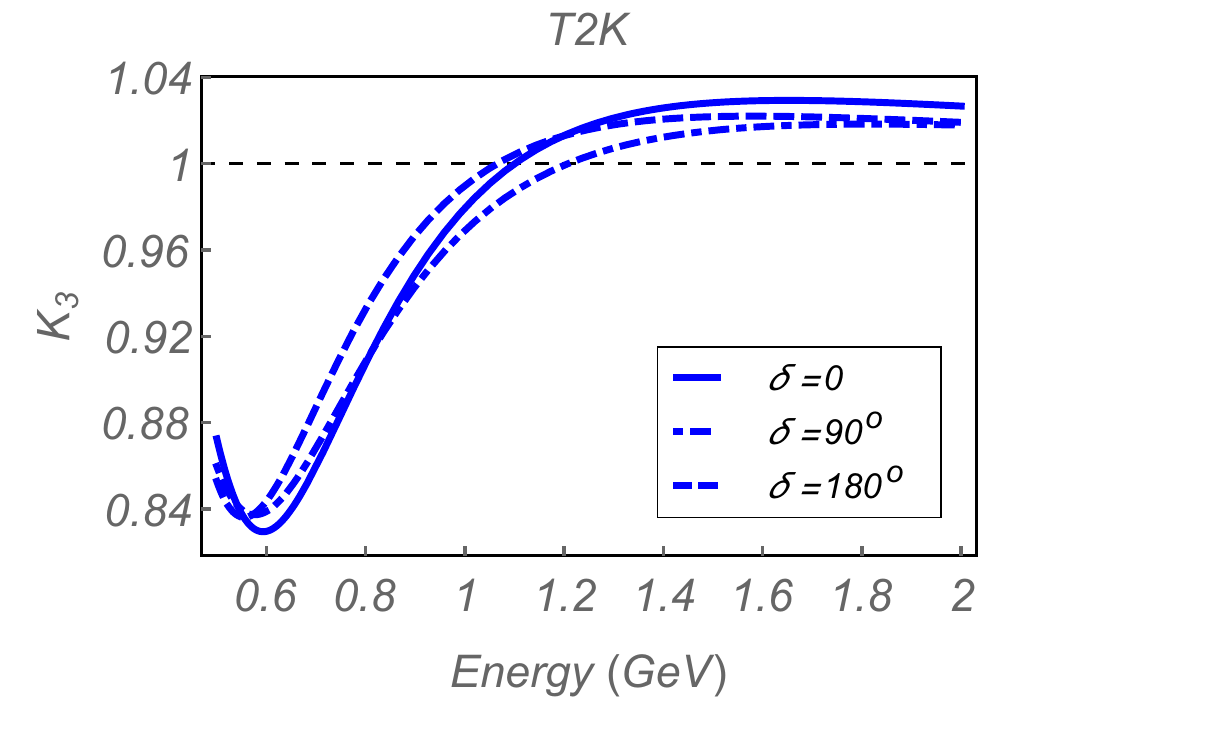}
		\end{tabular}
		\caption{Leggett-Garg function $K_3$ plotted against energy for DUNE (left), NO$\nu$A (middle) and T$2$K (right) experimental set-ups for different values of the $CP$ violating phase $\delta$. The initial neutrino state is taken as $\nu_\mu$ and the sign of $\Delta_{31}$ is positive. The time can be identified with the length which is 1300 km, 810 km and 295 km for DUNE NO$\nu$A and T2K, respectively.}
		\label{K3}
	\end{figure}
	\end{widetext}
\onecolumngrid
\begin{widetext}
	\begin{figure}[ht]
		\centering
		\begin{tabular}{ccc}
			\includegraphics[width=60mm]{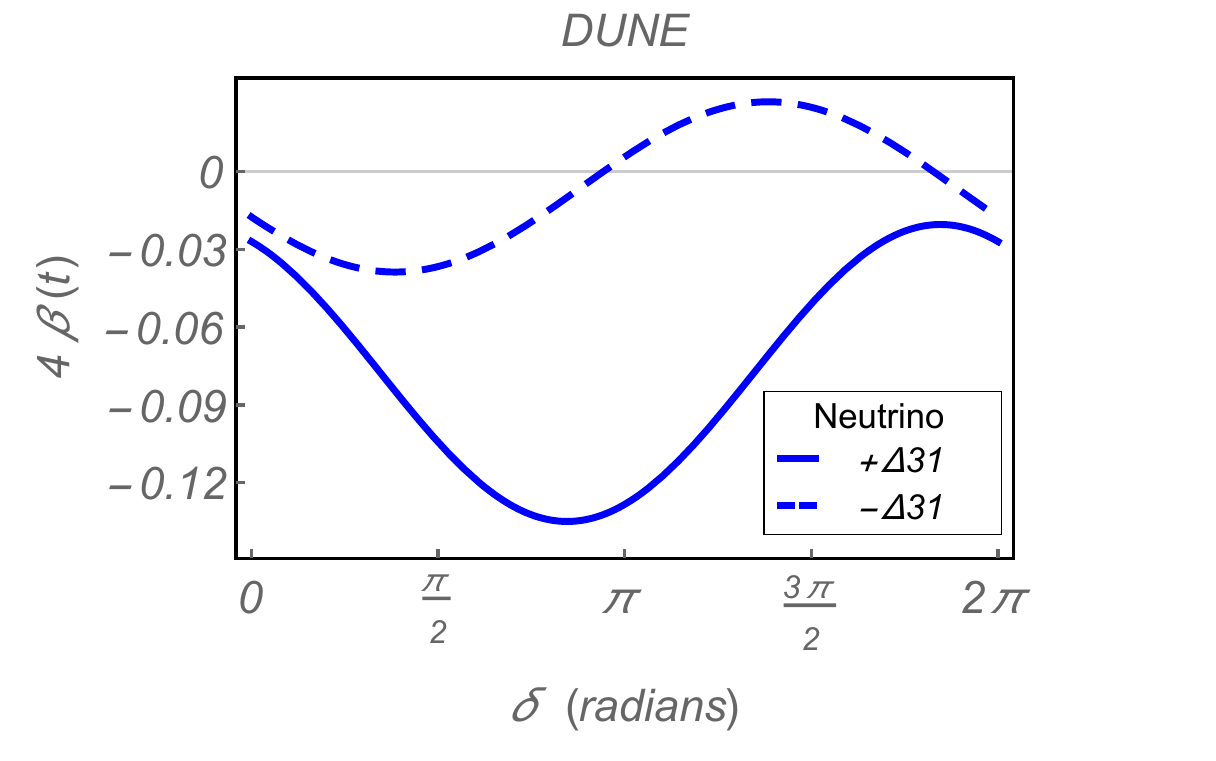}
			\includegraphics[width=60mm]{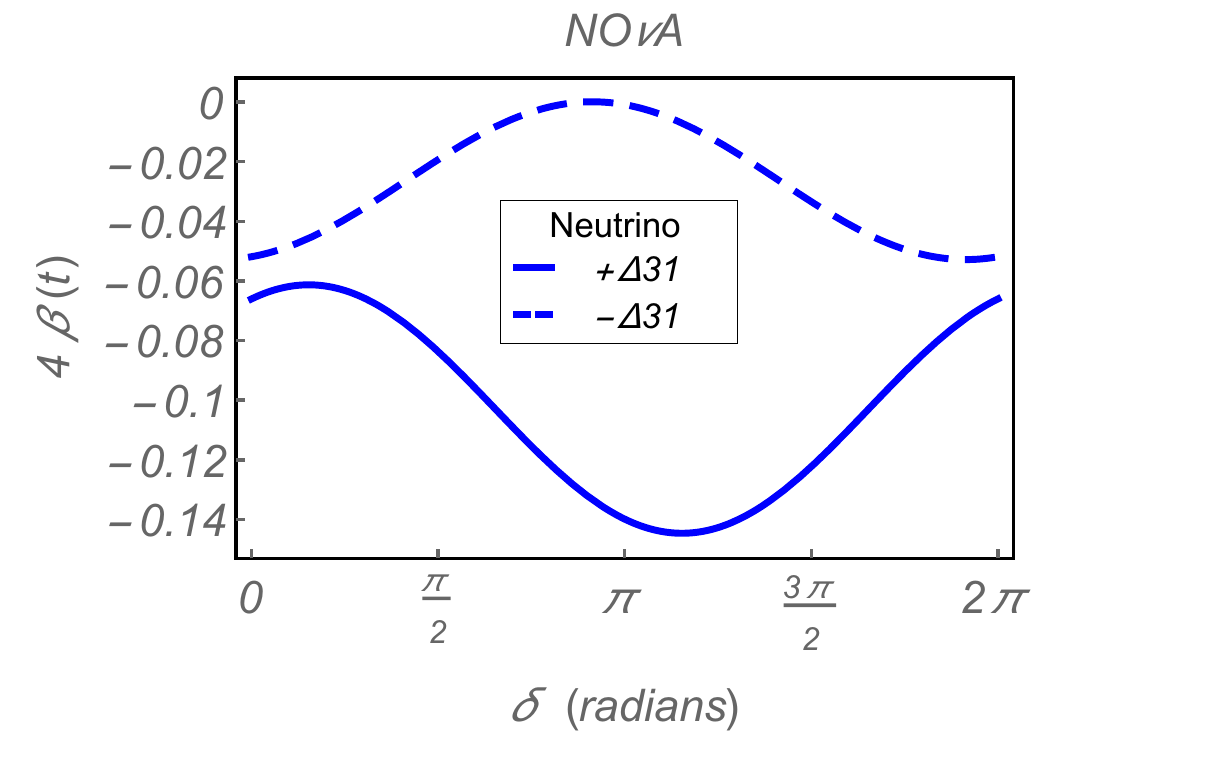}
			\includegraphics[width=60mm]{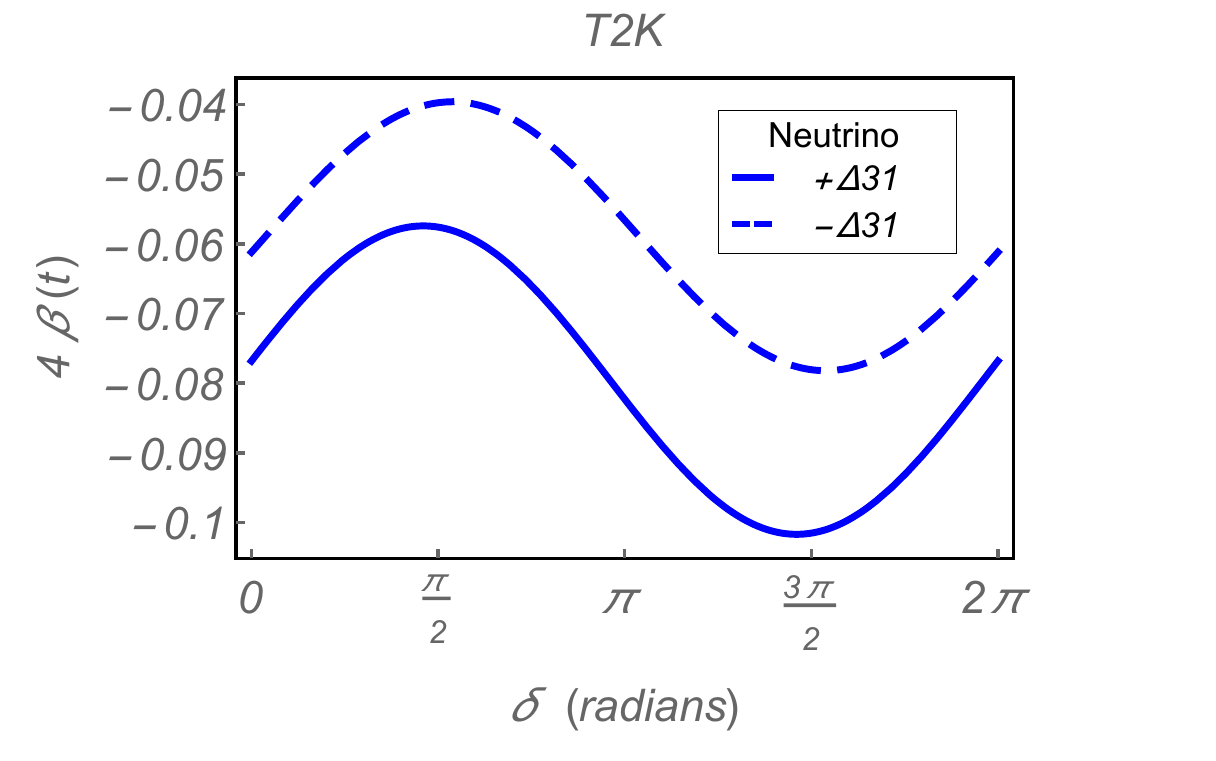}
		\end{tabular}
		\caption{ Plots of the term $4\beta(t)$, Eq. (\ref{mu-e}), with respect to $CP$ violating phase $\delta$ for DUNE (left), NO$\nu$A(middle) and $T2K$(right). Solid and dashed curves correspond to the positive and negative signs of  $\Delta_{31}$, respectively. The energies used are 3.5 GeV (DUNE), 2.5 GeV (NO$\nu$A) and 1.4 GeV(T2K) and pertain  to the maximum neutrino flux in the respective experimental setups.}
		\label{beta}
	\end{figure}
\end{widetext}

\begin{figure}[ht]
	\centering
	\begin{tabular}{ccc}
		\includegraphics[width=85mm,height=60mm]{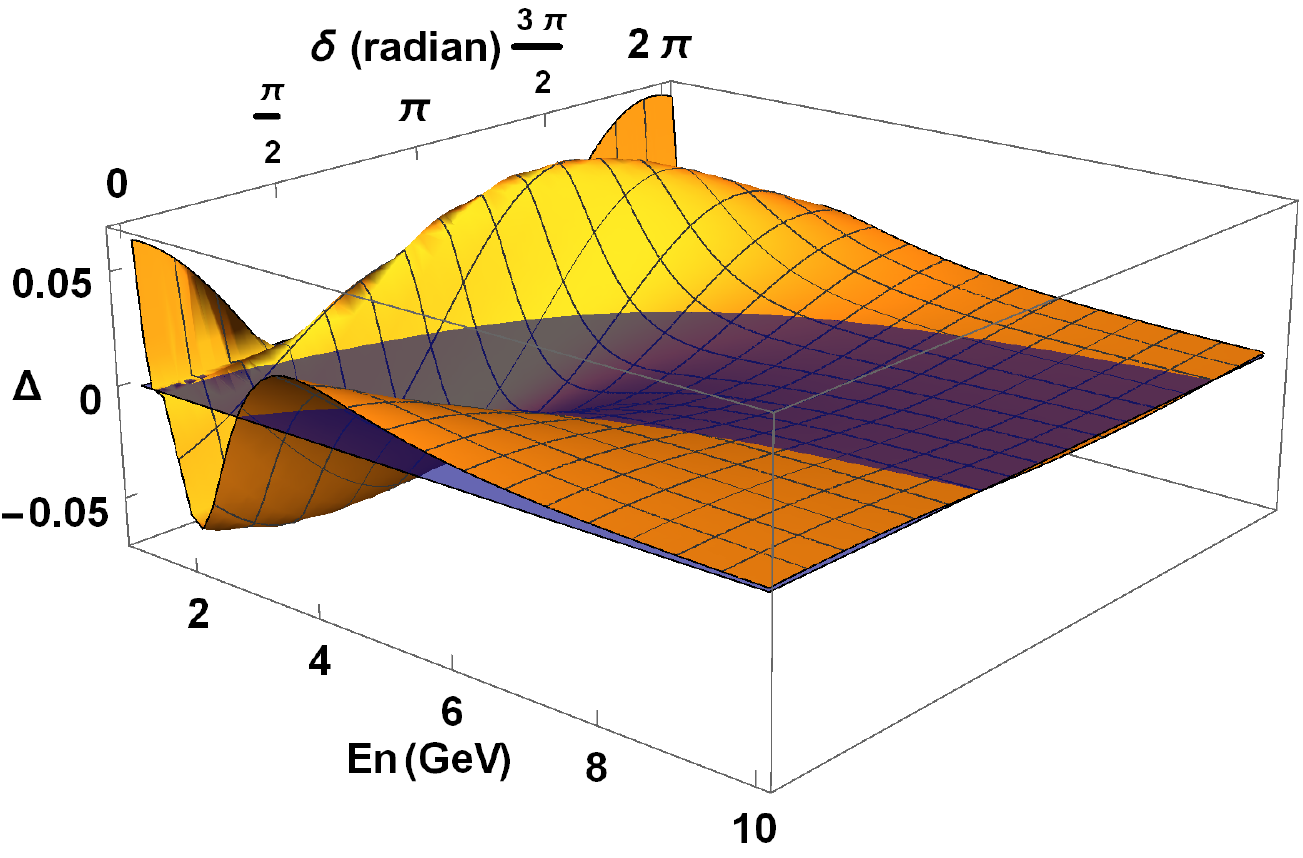}
	\end{tabular}
	\caption{Showing the difference $\Delta = K_3 - \tilde{K}_3$ as a function of energy  $E_n$ and $CP$ violating phase $\delta$. The function $\tilde{K}_3$ and $K_3$ are given by Eqs. (\ref{LGT}) and (\ref{mu-e}), respectively. The difference converges to zero for higher energies for all values of $\delta$, implying that one can safely approximate LGI by LGtI for high energy scenarios. This is because  the non-measurable terms $\alpha$ and $\beta$ in LGI can be approximated by $\frac{1}{2}$ (average value) and 0 (high energy limit), respectively, leading to LGtI. }
	\label{diff}
\end{figure}

	\subsection{Leggett-Garg inequality for neutrinos}

In this subsection we characterize the LG parameter $K_3$ in the case of three-flavour neutrinos and study the validity of LGI using input parameters from two of the current major experimental platforms, namely NO$\nu$A and T$2$K and the future experiment DUNE. Therefore, we focus on having a specific initial flavour eigenstate, i.e. $\nu_\mu$, and choose equal time intervals ($t_0=0,t_1=t,t_2=2t$). Herewith, our LG parameter $K_3$ becomes the sum of the following correlation functions
\begin{eqnarray}
K_3=C(0,t)+C(t,2t)-C(0,2t)\leq 1\;.
\end{eqnarray}
To compute the two-time correlation functions $C$ we need to employ the dichotomic observable $ \hat{Q} = 2 \ket{\nu_\alpha} \bra{\nu_\alpha} - \mathbb{1}$, which physically corresponds to asking whether the neutrino is still in the state $\ket{\nu_\mu}$ (associated outcome $1$) or has undergone a transition to another flavour state $\ket{\nu_\alpha}$ with $\alpha\not=\mu$ (associated outcome $-1$).
Straightforwardly one finds for	
\begin{eqnarray}
	C(0,t) &=& 4 \delta_{\alpha\mu} \langle \nu_{\mu}(t)|\nu_{\alpha} \rangle \langle \nu_\alpha | \nu_\mu (t) \rangle \nonumber \\ &-& 2 \langle \nu_\mu(t)| \nu_\alpha \rangle \langle \nu_\alpha|\nu_\mu(t) \rangle
	- 2  \delta_{\alpha\mu}   + 1\nonumber\\
&=&\left\lbrace\begin{array}{c} 2 \mathcal{P_{\mu \rightarrow \mu}}(t)-1\qquad\textrm{for}\;\alpha=\mu\\\\
1-2 \mathcal{P_{\mu \rightarrow \alpha}}(t) \qquad\textrm{for}\;\alpha\not=\mu
\end{array}\right. \label{C0tdef}
	\end{eqnarray}
where $\mathcal{P_{\mu \rightarrow \mu}}(t)$ is the surviving probability and $\mathcal{P_{\mu \rightarrow \alpha}}(t)$ is the transition probability.  Use has been made of the fact that the completeness condition in three flavour neutrino oscillation is $ \sum_{\alpha=e,\mu,\tau}^{} \ket{\nu_\alpha}\bra{\nu_\alpha} = \mathbb{1}$, leading to $P_{\mu e}(t) + P_{\mu \mu}(t) + P_{\mu \tau}(t) = 1$ in Eq. (\ref{C0tdef}).

The probabilities  with matter effect, depend on the neutrino energy E, the mass square differences $\Delta_{ij} =  m_j^2 - m_i^2$, the matter density parameter $A$, the mixing angles $\theta_{ij}$ and the CP violating phase $\delta$, i.e.,  $\mathcal{P_{\mu \rightarrow \alpha}} =  \mathcal{P_{\mu \rightarrow \alpha}}(E, t, A, \Delta_{12}, \Delta_{31}, \theta_{12}, \theta_{23}, \theta_{13}, \delta )$. For brevity in nomenclature, we suppress all the other dependencies except the time dependence.
The ongoing neutrino experiments NO$\nu$A and T2K studying the transition probabilities, $P_{ \mu \rightarrow e}(t)$. Thus we focus on the choice $\alpha=e$ in the following.

The tricky part is the correlation $C(t,2t)$ since it cannot be re-written into surviving or/and transition probabilities if one does not invoke the stationary condition, i.e. considering Legett-Garg type inequalities which has been done in details in \cite{Javid}. The correlation function computes to
\begin{align}
C(t,2t)&= 1 -2 P_{ \mu \rightarrow e}(t) -2 P_{ \mu \rightarrow e}(2t) \nonumber \\&+ 4\alpha(t) P_{ \mu \rightarrow e}(2t) + 4 \beta(t).
\end{align}
	
Finally, our LG function is given by
 \begin{align}\label{mu-e}
    K_3 &= 1 - 4 P_{ \mu \rightarrow e}(t) + 4 \alpha(t) P_{ \mu \rightarrow e}(2t) + 4 \beta(t),  
   \end{align}
with
\begin{align}\label{alpha-def}
\alpha(t) &= |U^{11}_f(t)|^2,
\end{align}
\begin{align}\label{beta-def}
\beta(t) &= Re\bigg[ U^{11}_{f}(t) \bar{U}^{21}_{f}(t) U^{22}_{f}(2t) \bar{U}^{12}_{f}(2t) \nonumber \\&+ U^{11}_{f}(t) \bar{U}^{31}_{f}(t) U^{32}_{f}(2t) \bar{U}^{12}_{f}(2t)\bigg].
\end{align}
Here $\bar{U}^{ij}_{f}={U}^{ij*}_{f}$ represents the complex conjugate of $U^{ij}_{f}$, the $ij$-th element of the flavour evolution matrix $U_f$ defined in Eq.~(\ref{ufvacuum}). 
It should be noticed that for $\alpha = 0.5$ and $\beta = 0$, we recover the \textit{stationarity} limit of LGI given by Eq. (\ref{LGT}).  An important observation is that for higher energies, the interference term $\beta$ converges to zero. Also, the term $\alpha$, which varies between zero and one, averages to $\frac{1}{2}$, thereby taking LGI to LGtI.  Therefore, LGtI can be thought of as a kind of LGI for higher neutrino energies.

	\subsection{Leggett-Garg inequality in neutrino experiments}

 Neutrino oscillation experiments are typically in the ultrarelativistic limit, and hence $t$ can be approximated by  $L$ \cite{Kim}. The distance traveled by the neutrinos in a given experiment is an important parameter and is called the baseline of the experiment. In this work, we have studied the LGI for two ongoing experiments NO$\nu$A \cite{NOvA1, NOvA2} and T2K  \cite{T2K1, T2K2} and the future experiment DUNE \cite{Dune}. The baseline for these three experiments is 1300 km for DUNE, 810 km for NO$\nu$A and 295 km for T2K. Both DUNE and NO$\nu$A  use the neutrinos from Fermi Lab with the energy between 2-10 GeV for former and 1-5GeV for the later. T2K uses the neutrino source from JPARC (Japan Proton Accelerator Research Complex) in Tokai with the approximate energy 0.5-1GeV. All these experiments use $\nu_\mu$ source and the neutrinos travel the matter density of about 2.8 gm/{$cm^3$} which corresponds to the density parameter $A\approx 1.01 \times 10^{-13}.$

In Fig.~(\ref{K3}) the LG function $K_3$ is depicted with respect to the energy corresponding to these three experimental setups for different $\delta$ values. It can be observed from Eq.~(\ref{mu-e}) that due to the presence of the $\alpha(t)$ and $\beta(t)$ terms, $K_3$ cannot be expressed only in terms of the neutrino survival and transition probabilities. This term $\beta(t)$ is plotted in  Fig.~(\ref{beta}) with DUNE,  NO$\nu$A  and $T2K$ parameters as a function of the $CP$-violating parameter $\delta$. Let us note that the $\beta(t)$ term is negative for the experimental parameters thus reducing the $K_3$ value, consequently a possible violation of the bound one. In general, it can be positive as in the case of  DUNE experiment. 
  It turns out that at higher energies, the difference between the LGtI and LGI converges to zero as depicted in Fig. (\ref{diff}). Therefore at higher energies, one can safely approximate LGI by LGtI. One can attribute this to the fact that at higher energies the $\beta$-term goes to zero. Since $0 \le \alpha \le 1$, which averages to $\frac{1}{2}$; under these approximations, i.e., $\alpha \approx \frac{1}{2}$ and $\beta \approx 0$, LGI reduces to LGtI.

\section{NIM condition in neutrino experiments}
	The NIM (non-invasive measureability) condition renders LGI difficult from the experimental point of view. This feature is captured in Eq. (\ref{mu-e}) in the form of non-measurable terms $\alpha$ and $\beta$, defined in Eqs. (\ref{alpha-def}) and (\ref{beta-def}), respectively. The NIM condition is intertwined with the LGI as a memory, or to be more precise, the lack of it, that the system has been \textit{measured}. The concept of \textit{stationarity} introduced in \cite{Huelga1995,Huelga1996} provides a way to bypass the problem. The two time correlations are functions of the time difference $t_j - t_i$ and one can replace the term $C(t, 2t)$, the source of $\alpha$ and $\beta$, by $C(0,t)$ leading to the LG type inequality Eq. (\ref{LGT}) which is completely expressed  in terms of the measurable neutrino oscillation probabilities. In our earlier work \cite{Javid}, we used this formalism to address the mass hierarchy problem in neutrino physics. In the present case, one can recover the \textit{stationarity} limit of LGI under the conditions that $\alpha=0.5$ and $\beta = 0$.  Since, $0 \le \alpha \le 1$, we can think of $\alpha = 0.5 $ as the average value. Also, the term $\beta \rightarrow 0$ for higher energies. Therefore, LGtI comes out as an average of LGI in high energy regime.

	\section{Conclusion}

	In this work we develop LG inequalities in the context of three flavour neutrino oscillations including matter as well as $CP$ violating effects. It turns out that the LG function contains non-measurable terms, $\alpha$ and $\beta$, apart from the experimentally measurable probabilities. Under the approximations $\alpha \approx 0.5$ and $\beta \approx 0$, one recovers the \textit{stationarity} limit of the LGI. These approximations  hold well in high energy experiments, since the interference term $\beta \rightarrow 0$ at higher energies, and $\alpha$ which varies between 0 and 1, can be approximated by its average value $\frac{1}{2}$. Therefore, the LGtI comes out as a high energy limit of LGI in three flavour scenario of neutrino oscillation.  Energy and baseline are seen to be the most important factors contributing to the violation.

	\section*{Acknowledgments}
     We acknowledge useful discussions with B.C. Hiesmayr  and G. Guarnieri. This work is partially supported by DST India-BMWfW Austria Project Based Personnel Exchange Programme for 2017-2018. SB acknowledges  partial financial support from Project No. 03 (1369)/16/EMR-II, funded by the Councel of Scientific \& Industrial Research, New Delhi.


\end{document}